\PassOptionsToPackage{pdfa}{hyperref}
\documentclass[sigconf, nonacm]{acmart}
\usepackage[a-2b]{pdfx}

\newcommand\vldbdoi{XX.XX/XXX.XX}
\newcommand\vldbpages{XXX-XXX}
\newcommand\vldbvolume{19}
\newcommand\vldbissue{11}
\newcommand\vldbyear{2026}
\newcommand\vldbauthors{\authors}
\newcommand\vldbtitle{\shorttitle} 
\newcommand\vldbavailabilityurl{URL_TO_YOUR_ARTIFACTS}
\newcommand\vldbpagestyle{empty}

\newcommand{\cheng}{\color{black}}
\newcommand{\jifan}{\color{black}}
\newcommand{\jifanb}{\color{black}}
\newcommand{\jianyang}{\color{black}}
\newcommand{\james}{\color{black}}
\newcommand{\chengb}{\color{black}}
\newcommand{\chengc}{\color{black}}
\newcommand{\chengd}{\color{black}}

\newcommand{\jifanrevision}{\color{black}}

\newcommand{\chengr}{\color{black}}
\newcommand{\chengrb}{\color{black}}

\newcommand{\jifancamera}{\color{black}}
\newcommand{\chengcr}{\color{black}}

\usepackage{makecell}
\usepackage{autobreak}
\usepackage[ruled, lined, linesnumbered]{algorithm2e}
\usepackage{soul}
\usepackage{balance}
\usepackage[many]{tcolorbox}

\newtcolorbox{reviewbox}[2][]{%
  enhanced,
  breakable,
  arc=0mm,
  boxrule=0.8pt,
  colback=white,
  colframe=black,
  coltitle=black,
  fonttitle=\bfseries,
  boxed title style={size=normal,colback=white,colframe=white,boxrule=0pt},
  attach boxed title to top center={yshift=-\tcboxedtitleheight/2},
  title={#2},
  #1}

\begin{document}
\sloppy



\title{GPU-Native Approximate Nearest Neighbor Search with IVF-RaBitQ: Fast Index Build and Search}








\author{Jifan Shi}
\affiliation{%
  \institution{Nanyang Technological University}
  \country{Singapore}
}
\email{jifan002@e.ntu.edu.sg}

\author{Jianyang Gao}
\affiliation{%
  \institution{Nanyang Technological University}
  \country{Singapore}
}
\email{gaoj0017@e.ntu.edu.sg}

\author{James Xia}
\affiliation{%
  \institution{NVIDIA}
  \country{USA}
}
\email{jamxia@nvidia.com}

\author{Tam\'{a}s B\'{e}la Feh\'{e}r}
\affiliation{%
  \institution{NVIDIA}
  \country{Germany}}

\email{tfeher@nvidia.com}

\author{Cheng Long}
\affiliation{%
  \institution{Nanyang Technological University}
  \country{Singapore}
}
\email{c.long@ntu.edu.sg}

\begin{abstract}
Approximate nearest neighbor search (ANNS) on GPUs is gaining increasing popularity for modern retrieval and recommendation workloads that operate over {\chengc massive} high-dimensional vectors. 
{\chengc Graph-based indexes deliver high recall and throughput but incur heavy build-time and storage costs. In contrast, cluster-based methods build and scale efficiently yet often need many probes for high recall, straining memory bandwidth and compute.}
Aiming to simultaneously achieve fast index build, high-throughput search, high recall, and low storage requirement for GPUs, we present IVF-RaBitQ (GPU), a GPU-native ANNS solution that integrates 
{\chengc the cluster-based method IVF}
with {\cheng RaBitQ} 
quantization 
into an efficient GPU {\chengb index} build/search pipeline.
Specifically, for index {\chengb build}, we develop a scalable GPU-native RaBitQ quantization method that enables fast and accurate low-bit encoding at scale.
For 
{\chengc search}, we develop GPU-native distance computation schemes for {\cheng RaBitQ}
codes and a fused search kernel to achieve high throughput with high recall.
{\chengc With IVF-RaBitQ implemented and integrated into the NVIDIA cuVS Library, experiments on cuVS Bench across multiple datasets show that IVF-RaBitQ offers a strong performance frontier in recall, throughput, index build time, and storage footprint.}
{\jifan For Recall$\approx$0.95, IVF-RaBitQ achieves {\jifancamera 3.0$\times$} higher QPS than the state-of-the-art graph-based method CAGRA, while also constructing indices {\jifancamera 14.7$\times$} faster on average. Compared to the cluster-based {\chengc method} IVF-PQ, IVF-RaBitQ delivers on average over {\jifancamera4.5$\times$} higher throughput while avoiding accessing the raw vectors for reranking.}

\end{abstract}

\maketitle

\pagestyle{\vldbpagestyle}
\begingroup\small\noindent\raggedright\textbf{PVLDB Reference Format:}\\
\vldbauthors. \vldbtitle. PVLDB, \vldbvolume(\vldbissue): \vldbpages, \vldbyear.\\
\href{https://doi.org/\vldbdoi}{doi:\vldbdoi}
\endgroup
\begingroup
\renewcommand\thefootnote{}\footnote{\noindent
This work is licensed under the Creative Commons BY-NC-ND 4.0 International License. Visit \url{https://creativecommons.org/licenses/by-nc-nd/4.0/} to view a copy of this license. For any use beyond those covered by this license, obtain permission by emailing \href{mailto:info@vldb.org}{info@vldb.org}. Copyright is held by the owner/author(s). Publication rights licensed to the VLDB Endowment. \\
\raggedright Proceedings of the VLDB Endowment, Vol. \vldbvolume, No. \vldbissue\ %
ISSN 2150-8097. \\
\href{https://doi.org/\vldbdoi}{doi:\vldbdoi} \\
}\addtocounter{footnote}{-1}\endgroup

\ifdefempty{\vldbavailabilityurl}{}{
\vspace{.3cm}
\begingroup\small\noindent\raggedright\textbf{Artifact Availability:}\\
The source code, data, and/or other artifacts have been made available at \url{https://github.com/Stardust-SJF/cuvs_rabitq}.
\endgroup
}

\section{Introduction}
\label{sec:introduction}

Approximate Nearest Neighbor Search (ANNS) {\cheng on high-dimensional vector data} has become a fundamental component in modern data management and machine learning. It supports key applications such as semantic search~\cite{10.1145/3447548.3467081}, Retrieval-Augmented Generation (RAG)~\cite{10.1145/3616855.3638207}, and recommender systems~\cite{8681160, 10.1145/3534678.3539071, 10.14778/3750601.3750700}. With the widespread adoption of embedding-based models~\cite{devlin2019bertpretrainingdeepbidirectional, radford2021learningtransferablevisualmodels, openai2024embedding, google2024embeddings}, production workloads now often 
{\cheng involve massive} high-dimensional vectors~\cite{41694, wiki:pageviewstats}. As a result, designing scalable and efficient ANNS solutions has become a long-standing and active research topic.

{\cheng A} variety of ANNS solutions have been proposed,
{\cheng among which graph-based methods and cluster-based methods are most commonly used in industrial systems}. 
Graph-based methods usually construct proximity graphs over data points and perform {\chengb greedy} 
traversal to locate nearest neighbors.
Representative algorithms, such as HNSW~\cite{10.1109/TPAMI.2018.2889473}, NSG~\cite{10.14778/3303753.3303754}, and Vamana graph~\cite{NEURIPS2019_09853c7f}, are known for their competitive recall-performance trade-offs.
Cluster-based methods, most notably inverted-file (IVF) indices, partition the vector space into coarse clusters and restrict search to a subset of clusters.
{\jianyang These methods offer simple data organization and efficient index construction, which enable easy integration with existing database systems.}
{\cheng Furthermore, both types of methods are often enhanced by leveraging vector quantization for reducing the costs of storing and accessing raw vectors and/or the costs of computing distances among raw vectors~\cite{NEURIPS2019_09853c7f, 5432202, 10.14778/3611479.3611537, gao2025the, 10.1145/3709730}.}
Most 
{\cheng of these solutions}
were originally developed and optimized for CPU architectures.

As data scale and query volume continue to grow, many real-world ANNS deployments choose to use GPUs as their accelerators 
due to GPUs' massive parallelism and high memory bandwidth.
{\chengrb For instance, cuVS, which is a GPU-based ANNS library from NVIDIA, has been widely used to handle batched queries on large-scale data~\cite{nvidia_cuvs_overview}.}
Existing GPU-based ANNS methods largely follow the same two paradigms as their CPU counterparts, namely graph-based approaches~\cite{9101583, 10.1145/3459637.3482344, 9739943, 10597683, 11045134, 10.5555/3768039.3768128, 10.1145/3736227.3736237, 9835618, cuhnsw} and cluster-based approaches~\cite{CHEN2019295, 10.5555/3724648.3724659, 7780592, 10.5555/3691825.3691827, 8733051, 10.1145/3749189}. 
Graph-based GPU methods redesign proximity graph building and traversal to better match GPU execution, leveraging warp-level parallelism, shared-memory caching, and fine-grained scheduling to mitigate control-flow divergence and irregular memory access. Representative systems such as SONG~\cite{9101583}, cuHNSW~\cite{cuhnsw}, GGNN~\cite{9739943}, and CAGRA~\cite{10597683} demonstrate that graph-based ANNS can achieve strong recall and competitive query performance on GPUs.

Compared with graph-based indices, cluster-based GPU methods offer a different set of trade-offs. 
A key merit is that cluster-based methods expose \emph{regular computation patterns} that align well with GPU primitives: coarse cluster selection can be implemented using dense linear algebra (e.g., GEMM for {\chengb computing query--centroid distances}), and the subsequent 
{\jianyang probing}
within selected clusters can be parallelized as large, contiguous streams of codes or vectors. 
This makes it natural to exploit GPU batching and achieve high throughput. Moreover, cluster-based indices typically have \emph{simpler structures} than proximity graphs, with more predictable memory access, which eases GPU memory management and improves memory coalescing. 

Motivated by the above considerations, this paper focuses on \emph{cluster-based} ANNS methods on GPUs. 
Within this family, IVF-Flat and IVF-PQ are widely used baselines~\cite{8733051, cuvs_github, 10.5555/3691825.3691827, 10.5555/3724648.3724659, 10.1145/3620665.3640360}. 
IVF-Flat stores full-precision vectors and performs brute-force scanning within selected clusters. Its accuracy is strong, but the scan stage is often dominated by memory bandwidth due to high-dimensional floating-point loads. 
{\jifanb IVF-PQ reduces memory traffic by compressing vectors using product quantization (PQ)~\cite{5432202}. However, it requires learning codebooks during index construction, introducing additional overhead on building complexity. 
Also, the compressed representation only provides approximate distance estimates without strict accuracy guarantees. As a result, achieving high recall often requires reranking with the raw vectors (e.g., raw vectors are stored in main memory and are accessed during search), which increases the memory footprint and complicates the overall pipeline.}
{\chengd Recently, a new vector quantization method called RaBitQ was proposed~\cite{10.1145/3654970,10.1145/3725413}, which improves upon PQ in several important aspects. First, RaBitQ does not require codebook training. Second, RaBitQ enables more accurate distance estimation with tight, theoretically grounded error bounds.
 RaBitQ has been combined with IVF for CPU-based ANN search and has shown consistent improvements over PQ~\cite{10.1145/3654970,10.1145/3725413}. Nevertheless, bringing RaBitQ to GPUs and integrating it with IVF is non-trivial and introduces several algorithmic and system challenges.
Specifically, the quantization procedure of RaBitQ involves searching over rescaling factors and evaluating candidates, 
which can be expensive and control-flow heavy.
}
{\cheng When implemented on CPU, RaBitQ's} distance evaluation step benefits from wide SIMD vectorization and specialized bit-level routines (e.g., AVX-512-style packed/shuffle operations and fast bit-scan / popcount primitives) to accelerate asymmetric inner products and reduce per-dimension overhead~\cite{gao2025the}. 
On GPUs with SIMT (Single Instruction, Multiple Thread) execution models, the control-flow heavy quantization becomes warp-divergent, and the CPU SIMD-optimized distance kernels have no direct equivalent. A naive port either underutilizes the GPU or incurs excessive memory traffic and kernel overhead. Realizing RaBitQ effectively on GPUs therefore requires GPU-native {\jifan design}, including scalable quantization strategies, GPU-oriented distance-evaluation primitives, and careful integration with the IVF probe-and-scan pipeline.

{\chengc In this paper,}
we present {\cheng \textbf{IVF-RaBitQ (GPU)}}, {\jifancamera hereafter \textbf{IVF-RaBitQ},} {\chengc the first} GPU-native ANNS {\jifan solution} that tightly integrates IVF coarse partitioning with RaBitQ quantization,
{\chengc for achieving fast index build, high-throughput search, high recall, and compact storage simultaneously}. 
{\chengc We summarize our technical contributions to the development of IVF-RaBitQ as follows.
{\jifanrevision \underline{First}, during index build,
apart from building the IVF index, we reformulate RaBitQ quantization for GPU execution and quantize data vectors into compact RaBitQ codes.}
Specifically, we design a cluster-at-a-time, block-per-vector quantization pipeline.
Within the pipeline, we develop a two-phase parallel grid search algorithm for exploring the rescaling factors in RaBitQ quantization. 
{\chengr This design is motivated by the observation that the objective function over rescaling factors is approximately unimodal.
{\chengrb Specifically, it applies a coarse-to-fine narrowing strategy {\jifanrevision which is inspired by} ternary search.}} 
These techniques enable scalable and high-throughput RaBitQ quantization for massive data vectors on GPUs.
\underline{Second}, during search, {\jifanrevision we develop GPU-native distance estimation methods {\chengr based on RaBitQ quantization codes}
for scanning IVF lists.}
Specifically, we adopt a two-stage distance estimation strategy for finding nearest neighbors of a query, namely a 
filtering stage based on the 1-bit RaBitQ code (which consists of the most significant bit of each dimension of the code), followed by a refinement stage using the ex-code (which consists of the remaining bits of each dimension of the code) only for candidates that pass the filter.
Since the filtering stage (which checks many data vectors) dominates the overall computation, 
we develop two GPU-native inner-product computation schemes, namely one based on \emph{lookup tables} and the other based on \emph{bitwise decomposition}, 
to accelerate this stage.
\underline{Third},}
we design a fused cluster-local search kernel {\chengc for finding nearest neighbors locally within a cluster, which} integrates distance filtering, refinement, and top-$K$ selection into a single GPU kernel, substantially reducing kernel-launch overhead and global-memory traffic.
{\chengc \underline{Fourth},
to support the index build and search pipelines, we design}
a GPU-oriented index layout that aligns data organization with access patterns of the fused kernel, enabling high memory coalescing and cache efficiency.

We also integrate {\cheng IVF-RaBitQ} build and search pipelines into {\jifanb NVIDIA \textbf{cuVS} library~\cite{cuvs_github} and \textbf{cuVS Bench}~\cite{rapids_cuvs_bench}}, enabling reproducible benchmarking and direct comparisons with existing GPU baselines under a unified evaluation framework.
{\chengc We conduct extensive experiments using cuVS Bench with diverse datasets,
}
demonstrating IVF-RaBitQ's advantages in the trade-off among recall, throughput, index build time, and storage footprint.
{\jifanb For query performance, we evaluate methods {\chengcr mainly} under large-batch query settings to fully utilize GPU resources.} Compared with the state-of-the-art graph-based {\chengc GPU} method, CAGRA, IVF-RaBitQ achieves comparable or higher throughput {\jifancamera (0.8$\times$\textasciitilde8.2$\times$ speedup for Recall$\approx$0.95)} while reducing index construction time by {\jifanb up to} order of magnitude {\jifancamera (3.4$\times$\textasciitilde55.3$\times$ speedup)} with far less storage requirement.
Compared with IVF-PQ implemented in NVIDIA cuVS Library, {\cheng IVF-RaBitQ} achieves significantly higher query throughput {\jifancamera (1.4$\times$\textasciitilde58.3$\times$ speedup for Recall$\approx$0.95)} and accuracy under similar index build cost and storage budget. These results 
show that IVF-RaBitQ is a practical and scalable solution for large-scale GPU-based ANNS problems.


\section{Preliminaries}
\label{sec:pre}

\subsection{Approximate NN Search and the IVF Index}
 {\cheng In this paper, we {\chengc target} the \emph{batched} setting of approximate nearest neighbor search (ANNS): given a batch of queries} $\mathbf{Q}\in\mathbb{R}^{n_q\times D}$, 
 {\chengc we want to find}
 the {\chengb $K$} approximate nearest neighbors from a set of $N$ data vectors $\mathbf{O} \subseteq \mathbb{R}^D$ for every query in the batch.


In the IVF index, 
vectors are first partitioned into clusters using a clustering algorithm (e.g., K-means~\cite{mcqueen1967some} or its variations~\cite{10.5555/1283383.1283494, 10.1007/978-3-662-44415-3_4}). 
{\chengcr The search process} is usually carried out in two stages~\cite{5432202}.
First, 
{\jianyang {\james given a query vector}, a few nearest clusters are identified based on the distances between their centroids to the query.}
Then, 
these clusters are probed by computing the distances between the vectors in the clusters and the query vector and finding the nearest neighbors.
With different settings of the number of clusters to probe, IVF provides different accuracy-efficiency trade-offs.




\subsection{{\cheng The RaBitQ Quantization Method}}
\label{subsec:rabitq}
{\cheng We 
{\chengb first review the} process of quantizing a raw vector $\mathbf{o}_\mathrm{r}$
{\chengb in RaBitQ~\cite{10.1145/3654970,10.1145/3725413}}. 
First, it normalizes $\mathbf{o}_\mathrm{r}$ to a unit vector $\mathbf{o}$ wrt a center vector $\mathbf{c}$, i.e., we have $\mathbf{o} = \frac{\mathbf{o}_\mathrm{r} - \mathbf{c}}{\|\mathbf{o}_\mathrm{r} - \mathbf{c}\|}$. 
Then, it focuses on quantizing the unit vector $\mathbf{o}$. 
{\chengcr Note that the direction of $\mathbf{o}_\mathrm{r}$ is captured by $\mathbf{o}$ and
the norm of $\mathbf{o}_\mathrm{r}$ can be pre-computed and stored.}
Specifically,
it constructs a codebook by shifting, normalizing, and randomly rotating vectors whose coordinates are $B$-bit unsigned integers, where $B$ could be 1, 2, 3, .... 
Formally, given the number of dimensions of a vector $D$ and the number of bits used to quantize each dimension $B$}, 
the codebook $\mathcal{G}_\mathrm{r}$ is defined as follows:

\begin{equation} \label{eq:G}
\mathcal{G} := \left\{ -\frac{2^B - 1}{2} + u \Biggm| u = 0, 1, 2, 3, \dots, 2^B - 1 \right\}^D
\end{equation}
\begin{equation} \label{eq:Gr}
\mathcal{G}_\mathrm{r} := \left\{ \mathbf{P} \frac{\mathbf{x}}{\|\mathbf{x}\|} \Biggm| \mathbf{x} \in \mathcal{G} \right\}
\end{equation}
{\cheng where $\mathbf{P}$ is a random orthogonal matrix~\cite{johnson1984extensions}.}

{\cheng To quantize the vector $\mathbf{o}$, it}
takes $\mathbf{o}$'s nearest vector in the codebook $\mathcal{G}_\mathrm{r}$, {\cheng which we denote by $\mathbf{\hat{o}}$}, as its quantized vector. Let {\chengb $\hat{\mathbf{x}}$} be the vector in $\mathcal{G}$, {\cheng which corresponds to $\hat{\mathbf{o}}$ (note that each vector in $\mathcal{G}_\mathrm{r}$ 
corresponds to a vector in $\mathcal{G}$ (before {\chengc normalization and} rotation)). 
This process can be formally described as follows.}

\begin{align}
\hat{\mathbf{x}} &= \underset{\mathbf{x} \in \mathcal{G}}{\operatorname{arg\,min}} \left\| \mathbf{P} \frac{\mathbf{x}}{\|\mathbf{x}\|} - \mathbf{o} \right\|^2 
= \underset{\mathbf{x} \in \mathcal{G}}{\operatorname{arg\,max}} \left\langle \frac{\mathbf{x}}{\|\mathbf{x}\|}, \mathbf{o'} \right\rangle \label{eq:y_bar_max}
\end{align} 
{\cheng where $\mathbf{o}'=\mathbf{P}^{-1} \mathbf{o} $ is the data vector after rotation.}

{\cheng According to Equation~\ref{eq:y_bar_max}, the quantization process is to find $\hat{\mathbf{x}}$, which is the vector $\mathbf{x} \in \mathcal{G}$ that maximizes {\jianyang the cosine similairty} 
{\chengb with} the data vector after rotation, i.e., $\mathbf{o}'$.
{\chengb According to the analysis in~\cite{10.1145/3725413},} 
the vector {\chengb $\hat{\mathbf{x}}$} must be closest to a \textit{re-scaled} vector $\mathbf{o}'$ for some \textit{{\chengc rescaling} factor} $t$, i.e., $t \cdot \mathbf{o}'$. 
Therefore, while there are $2^{B \cdot D}$ {\chengb vectors} in $\mathcal{G}$, the quantization process does not need to enumerate all of them. Instead, it only enumerates those which are closest to $t\cdot \mathbf{o}'$ for some $t$. 
Furthermore, while there are infinite {\chengc rescaling} factors, there are only $D\cdot 2^{B-1}$ \textit{critical} ones, at which the vectors that are closest to $t \cdot \mathbf{o}'$ {\chengb in $\mathcal{G}$} would change~\cite{10.1145/3725413}. 
As a result, the quantization process can explore these \textit{critical} {\chengc rescaling} factors only. 
In summary, the quantization process is to identify and enumerate $D\cdot 2^{B-1}$ critical {\chengc rescaling} factors. For each {\chengb factor $t$}, it finds the vector $\mathbf{x}$ in $\mathcal{G}$ that is closest to $t\cdot \mathbf{o}'$ and computes {\chengb its cosine similarity with $\mathbf{o}'$}.
Finally, it returns the vector {\chengb $\hat{\mathbf{x}}$, which} has the greatest {\chengb cosine similarity}.}

{\cheng 
Given a raw query vector $\mathbf{q}_\mathrm{r}$, we review the process of estimating the Euclidean distance between $\mathbf{q}_\mathrm{r}$ and a raw data vector $\mathbf{o}_\mathrm{r}$ in RaBitQ as follows. It 
first rotates $\mathbf{q}_\mathrm{r}$ with the same matrix $\mathbf{P}$ used for rotating the data {\chengb vectors}. We denote the 
query vector after rotation by $\mathbf{q}'$. It then estimates the distance between $\mathbf{q}_\mathrm{r}$ and $\mathbf{o}_\mathrm{r}$ based on several variables: namely (1) the dot product between the RaBitQ code of $\mathbf{o}$, i.e., {\chengb $\hat{\mathbf{x}}$}, and the rotated query vector, $\mathbf{q}'$ and (2) some factors~\cite{gao2025the}. 
{\jianyang For detailed expressions of the computation, we refer readers to the documentation of the RaBitQ Library~\cite{rabitq_documentation}. }
}
\section{The GPU-based IVF-RaBitQ Method}
\label{sec:method}
\subsection{Overview}
\label{subsec:overview}

IVF-RaBitQ (GPU) is a GPU-native ANNS solution 
{\jifan that} combines an IVF coarse partition with RaBitQ 
quantization and implements both \textit{index build} and \textit{search} as GPU-friendly pipelines. The overview of its workflow is described as follows.


{\cheng \smallskip\noindent\textbf{Index build.}}
Given {\chengb a set of $N$ raw data vectors $\mathbf{O} \in \mathbb{R}^{N \times D}$ 
(
$D$ is the vector dimensionality)}, we first learn $n_k$ IVF centroids and assign each data vector to its nearest centroid, thereby forming $n_k$ inverted lists. 
{\chengd Within each cluster, we normalize the vectors with respect to the centroid, rotate them with a random orthogonal matrix $\mathbf{P}$, and quantize them using RaBitQ and obtain their codes.}
Each code is stored in two parts: (i) the most significant bit of each dimension ({\cheng which we call the \textit{1-bit RaBitQ code} or simply the \textit{1-bit code}}) and (ii) the remaining bits ({\cheng which we call the \textit{ex-code}}). 
To enable efficient large-scale RaBitQ quantization on GPU, we write GPU kernels to process vectors in a cluster-wise manner,
and develop GPU-friendly algorithms for exploring the rescaling {\cheng factors} used in RaBitQ quantization 
{\chengb (\underline{Section~\ref{subsec:index_build}})}.

{\cheng \smallskip\noindent\textbf{Search.}}
In IVF-RaBitQ, queries are processed in batches {\chengb (\underline{Section~\ref{subsec:search_pipeline}})}.
Given a query batch $\mathbf{Q}\in\mathbb{R}^{n_q\times D}$ and a probe count $n_\mathrm{probe}$, we first apply the same random orthogonal transformation $\mathbf{P}$ used in index {\chengb build to the queries} and select the $n_\mathrm{probe}$ closest IVF clusters for each query.
The search workload is then organized at the granularity of \textit{$(\text{query}, \text{cluster})$ pairs}, enabling independent and parallel processing within each probed cluster.
For each $(\text{query}, \text{cluster})$ pair, we perform a cluster-local search that evaluates candidates stored in the corresponding inverted list
{\cheng and finds the {\chengb top-$K$} candidates (i.e., $K$ nearest neighbors) of the query within the cluster. Specifically, it}
adopts a two-stage distance estimation strategy: a 
filtering stage based on the 1-bit RaBitQ code, followed by a refinement stage using the ex-code only for candidates that pass the filter.
Since the filtering stage {\chengc checks many data vectors and} dominates the overall computation, we design GPU-native inner-product computation methods to accelerate this stage ({\chengb \underline{Section~\ref{subsec:ip_gpu}}}), and further integrate it with filtering, refinement, and top-$K$ selection into a fused GPU kernel to reduce kernel-launch and memory-access overheads ({\chengb \underline{Section~\ref{subsec:fused_cluster_search_kernel}}}).
Finally, the top-$K$ candidates obtained from all probed clusters are merged to produce the global top-$K$ result for each query.

{\cheng \smallskip\noindent\textbf{GPU index layout.}}
{\jifanb To efficiently support ANNS on GPU, we design a flattened Compressed Sparse Row (CSR)-like layout \cite{10.1145/1583991.1584053}
{\chengcr for storing the clusters of the IVF index.}} 
{\chengcr We use it to store other elements of the index as well, including}
separate contiguous arrays for 1-bit codes, ex-codes, and auxiliary factors. 
We further adopt an interleaved layout for 1-bit codes that is tailored to the access pattern of the dominant filter stage, leading to improved memory coalescing on the GPU ({\chengb \underline{Section~\ref{subsec:optimized_GPU_data_layout}}}).

\subsection{Index Build on GPU}
\label{subsec:index_build}

\subsubsection{{\chengc Clustering, Normalization and Rotation}}
\label{subsec:kmeans_assign}

We run GPU-based Balanced K-means~\cite{cuvs_kmeans_cpp_api_2512} to learn $n_k$ centroids and assign each database vector to its nearest centroid. This produces the centroid matrix $\mathbf{C}\in\mathbb{R}^{n_k\times D}$ and a cluster id for each vector.
%
%
%
Within each cluster, we 
normalize {\chengc each vector} to unit $\ell_2$ norm {\chengc wrt its corresponding centroid}.
For simplicity, when the context is clear, we refer to the normalized {\chengc vectors} as (data) vectors and denote a data vector by $\mathbf{o}$.
%
%
We {\chengc then} apply a shared random orthogonal matrix $\mathbf{P}$ to rotate data vectors and centroids, which is implemented using General Matrix Multiplication (GEMM). {\jifancamera Alternatively, we implement the roatation as a fast Hadamard transform combined with Kac's random walk on GPU, {\chengcr which we borrow} from the RaBitQ Library~\cite{rabitq_documentation}.} Since $\mathbf{P}$ is orthogonal, {\chengc the rotation} preserves inner products and distances between data vectors, centroids, and therefore does not alter exact nearest-neighbor relationships.
{\cheng We denote a data vector after rotation by $\mathbf{o}'$.}

\subsubsection{GPU-based RaBitQ Quantization for Data Vectors}
\label{subsec:quant_gpu_rabitq}

Given a set of $N$ data vectors $\mathbf{o}$
and a budget of $B$ bits per dimension, we aim to develop a GPU-based solution for quantizing the data vectors using RaBitQ, i.e., computing the RaBitQ codes of the data vectors. 

\smallskip\noindent
\textbf{Kernel granularity and thread mapping.}
 {\jifan We first decide the kernel granularity (i.e., how many vectors are processed per launch), which directly affects the parallelism effectiveness by determining how computation and memory resources are scheduled.} Quantizing \textit{one vector at a time} is prohibitively expensive when the dataset contains millions to billions of vectors, because kernel-launch overhead and per-vector scheduling costs accumulate quickly, and the GPU’s massive parallelism (hundreds of thousands of concurrent threads) is difficult to fully utilize. At the other extreme, quantizing \textit{all vectors at once} can saturate the GPU, but it imposes a high GPU memory footprint (especially for intermediate buffers) and is inflexible across datasets of different sizes. We therefore adopt a \textit{cluster-at-a-time} design: each kernel launch processes the vectors in one cluster, which gives a natural and bounded unit of work aligned with the IVF structure, and makes the implementation largely data-size independent. 

Next, we decide the mapping between GPU execution units and vectors. A straightforward approach is to assign a single thread to each vector, analogous to the parallelization strategy typically employed on CPUs. However, this prevents intra-vector parallelization: the computation required by RaBitQ for a single vector (which involves exploring multiple critical scaling factors) becomes serialized within one thread, limiting throughput. Moreover, adjacent threads in a warp would typically access different vectors located at different regions in global memory, which can easily lead to \textit{uncoalesced} memory accesses and lower effective bandwidth. Additionally, under a cluster-at-a-time execution schedule, the number of concurrently processed vectors may be insufficient to fully utilize the GPU. Thus, we adopt a {\cheng \textit{block-per-vector}} mapping as a robust and flexible design point, which has several advantages: (1) it enables fine-grained parallelism for a vector, e.g., it can explore multiple critical scaling factors in parallel; and (2) the threads in the same warp access contiguous (or near-contiguous) locations of the same vector, yielding \textit{coalesced} global-memory accesses. This improves both compute efficiency and memory throughput, and makes it easier to maintain high occupancy when operating cluster by cluster. 



\smallskip\noindent
\textbf{Efficient exploration of {\chengc rescaling} factors.}
The standard RaBitQ quantization algorithm enumerates {\chengb $D \cdot 2^{B-1}$} critical {\chengc rescaling} factors in ascending order and evaluates them using incremental updates~\cite{10.1145/3725413}. 
To support this efficient enumeration, previous CPU implementation maintains per-vector state, typically using a small priority queue to generate the next {\chengc rescaling} factor without explicitly materializing or sorting all candidates~\cite{10.1145/3725413}.  
However, {\cheng this sequential enumeration relies on fine-grained state updates and involves irregular control flow, which is not friendly to GPU execution model}. 

To achieve high accuracy across dimensionalities while maintaining high performance, we propose a GPU-friendly search for the rescaling factor. 
{\cheng Recall that the quantization process of RaBitQ explores a set of critical {\chengc rescaling} factors. For each factor $t$, it identifies a vector $\mathbf{x} \in \mathcal{G}$ (which is the closest to $t\cdot \mathbf{o}'$) and computes the dot product $\left\langle \frac{\mathbf{x}}{\|\mathbf{x}\|}, \mathbf{o}' \right\rangle$ (or $\left\langle \mathbf{x}, \mathbf{o}' \right\rangle / \|\mathbf{x}\|$). Finally, it returns the vector {\chengb $\hat{\mathbf{x}}$, which has} the greatest dot product. 
This quantization process can be regarded as one of optimizing an objective function 
$\left\langle \mathbf{x}, \mathbf{o}' \right\rangle / \|\mathbf{x}\|$
of $t$.
We observe this objective function is often \textit{approximately unimodal} in practice (i.e., it tends to increase then decrease with a single peak (possibly with small local fluctuations)).}
A classical \textit{ternary search} locates the maximum of a unimodal function by repeatedly shrinking the interval; however, it is inherently iterative and not ideal for GPU control flow. 

Inspired by ternary search, we adopt a \textit{two-phase grid search} that achieves a similar coarse-to-fine narrowing effect using a fixed number of GPU-parallel rounds. Given a $D$-dimensional normalized vector $\mathbf{o}'$ {\cheng (after rotation)} and bit-width $B$, {\jifan we first determine the search range of factors that could affect quantization result}.
The algorithm then proceeds in two phases. In the coarse search phase, we uniformly sample $N_{\mathrm{coarse}}$ candidate {\chengc rescaling} factors across the search range and evaluate them in parallel using multiple threads. For each candidate $t$, we 
{\cheng find the code $\mathbf{x}_{\mathrm{cur}}$ that is closest to $t\cdot \mathbf{o}'$}
and compute the objective $\langle \mathbf{x}_{\mathrm{cur}}, \mathbf{o}' \rangle / \|\mathbf{x}_{\mathrm{cur}}\|$.
A parallel reduction then identifies the best candidate $t_{\mathrm{center}}$. In the fine search phase, we narrow the search range to a small neighborhood around $t_{\mathrm{center}}$ and repeat the parallel evaluation with $N_{\mathrm{fine}}$ samples to refine the solution. The final {\chengc rescaling} factor $t_{\mathrm{max}}$ is used to produce the quantization code $\hat{\mathbf{x}}$. On the GPU, for each vector, we assign a thread block to do the sampling and the reduction in parallel. 
{\jifanrevision In this way,
the {\chengrb algorithm} evaluates only $N_{\mathrm{coarse}}+N_{\mathrm{fine}}$ sampled factors per vector, rather than enumerating all $O(D \cdot 2^B)$ critical factors as in the standard RaBitQ quantization algorithm. 
}

\subsection{Search Pipeline on GPU}
\label{subsec:search_pipeline}

\subsubsection{Batch Rotation}
\label{subsec:batch_rotation}

Let $\mathbf{Q}\in\mathbb{R}^{n_q\times D}$ be a batch of queries.
We rotate the batch using the same shared orthogonal matrix $\mathbf{P}$ as in the index {\chengb build}. This produces a contiguous rotated query batch $\mathbf{Q'}$ that is reused by all subsequent stages in the search pipeline.



\subsubsection{Cluster Selection}
\label{subsec:cluster_selection}

We select the $n_\mathrm{probe}$ closest clusters for each query.
Let $\mathbf{C}\in\mathbb{R}^{n_k\times D}$ be the centroid matrix and $\mathbf{Q'}\in\mathbb{R}^{n_q\times D}$ be the rotated query batch.
The squared Euclidean distance matrix between queries and centroids can be written in matrix form as
\begin{equation}
\mathbf{D}
=
{\cheng \mathbf{s}_\mathrm{q}}\mathbf{1}_{n_k}^{\top}
+
\mathbf{1}_{n_q}\mathbf{s}_\mathrm{c}^{\top}
-
2\,\mathbf{Q'}\mathbf{C}^{\top},
\end{equation}
where ${\cheng\mathbf{s}_\mathrm{q}}\in\mathbb{R}^{n_q}$ and $\mathbf{s}_\mathrm{c}\in\mathbb{R}^{n_k}$ store the squared $\ell_2$ norms of queries and centroids, respectively.
The dominant term $\mathbf{Q'}\mathbf{C}^{\top}$ is computed via a single GEMM on the GPU, and a parallel top-$K$ kernel~\cite{10.1145/3581784.3607062} is then used to select the $n_\mathrm{probe}$ nearest clusters per query.

\subsubsection{Probe Scheduling via Workload Reorganization}
\label{subsec:probe_scheduling}

After cluster selection, we reorganize the workload explicitly as a set of $(\text{query},\text{cluster})$ pairs, one pair per probed cluster of each query. 
We sort these pairs by cluster identifier before launching the search kernels for large batches. This 
reorganization is adopted from cuVS’s IVF-PQ~\cite{cuvs_github}, which converts the irregular per-query probe lists into an execution order that is friendly to GPU: first, blocks that process the same (or nearby) clusters execute closer in time, improving cache locality for the RaBitQ codes; second, the clusters of a fixed query across the batch are less likely to be probed concurrently, which gives a tighter running top-$K$ threshold (i.e., the current $K$-th best distance for each query) to enable more aggressive pruning in later probes. 
Before these workloads are mapped to SMs on the GPU, we compute query-dependent factors once per query and prepare the query representation needed by the chosen inner-product method (Section~\ref{subsec:ip_gpu}); these results are reused across that query's $n_\mathrm{probe}$ cluster scans.

\subsubsection{Cluster-local Search and Two-stage Distance Estimation}
\label{subsec:cluster_local_search}

For each $(\text{query},\text{cluster})$ pair, we compute \textit{cluster-local} top-$K$ candidates {\cheng (which are the query's $K$ nearest neighbors {\chengb locally} within the cluster)} using a two-stage estimator:
\begin{itemize}
\item \textbf{Stage 1 (filter):} compute a fast 1-bit distance estimate ({\cheng and} a lower bound) {\chengc based on the 1-bit RaBitQ code} and prune vectors whose values exceed the current threshold (i.e., the current $K$-th best distance for each query). 
\item \textbf{Stage 2 (refinement):} for the surviving candidates, read additional bits {\chengc of the RaBitQ code} (ex-code) to compute a more accurate estimated distance and then select the cluster-local top-$K$ candidates.
\end{itemize}
For simplicity, we refer to the estimated distances in Stage~1 and Stage~2 as the \textit{1-bit estimated distance} and the \textit{refined estimated distance}, respectively. 
Since the inner-product computation between the query and the 1-bit RaBitQ codes in Stage~1 dominates the computation
{\cheng (note that the computation is conducted for all vectors in the probed clusters)}
~\cite{gao2025the}, we develop two GPU-native inner-product computation schemes tailored to the GPU execution model (\underline{Section~\ref{subsec:ip_gpu}}).
Moreover, this two-stage estimator is realized on the GPU as a four-step cluster-local search procedure, {\cheng including (1)} 1-bit estimated distance computation, {\cheng (2)} candidate selection (i.e., retaining vectors whose estimated distances pass the current threshold), {\cheng (3)} refined estimated distance computation, and {\cheng (4)} in-block top-$K$ selection (i.e., selecting the best $K$ candidates within a thread block). 
{\cheng These steps} are fused into a single GPU kernel to reduce kernel launch overhead and global-memory traffic
(\underline{Section~\ref{subsec:fused_cluster_search_kernel}}).

\subsubsection{Batch-level Merge}
\label{subsec:batch_merge}

Each block outputs cluster-local candidates. For each query, we merge candidates from its $n_\mathrm{probe}$ clusters and run a GPU top-$K$ to produce the final approximate nearest neighbours.

\subsection{Kernel and Data-Layout Co-design}
\label{subsec:co_design}

\subsubsection{GPU Inner-Product Computation for 1-bit Estimated Distance}
\label{subsec:ip_gpu}
%
{\cheng Recall in RaBitQ, we need to compute the inner product between a query vector $\mathbf{q'}$ (after rotation) and the 1-bit RaBitQ code {\chengd of a data vector}, {\cheng which we denote by $\mathbf{\hat{x}}_\mathrm{b}$,} for {\chengb computing the 1-bit estimated} distance between the query and the vector {\chengd (Section~\ref{subsec:rabitq})}.}
{\cheng That is,} the inner product is computed in an asymmetric form, where query vectors are represented in {\jifanb higher} precision {\jifanb (e.g., floating-point)},
while data vectors are encoded as compact 1-bit RaBitQ codes. 
Existing CPU-oriented acceleration techniques~\cite{gao2025the} {\chengc are} based on SIMD instructions {\chengc and} cannot be directly ported to GPUs due to fundamental differences in execution models and available instruction sets. Thus, a straightforward approach would be to expand the binary codes into byte values and reuse conventional inner-product operators. However, such a design is inefficient on GPUs, as it introduces unnecessary type conversions and leads to suboptimal utilization of GPU compute and memory resources.

To address the problem, we develop two GPU-native inner-product computation schemes for this type of inner product.
The first scheme adopts a {\cheng \textit{lookup table based strategy}} that trades shared-memory capacity for reduced arithmetic operations, while the second scheme reformulates the inner product into a {\cheng \textit{bitwise decomposition}} that exploits GPU-friendly bitwise operations.

\begin{figure*}[!thbp] 
    \centering
    \includegraphics[width=1\linewidth]{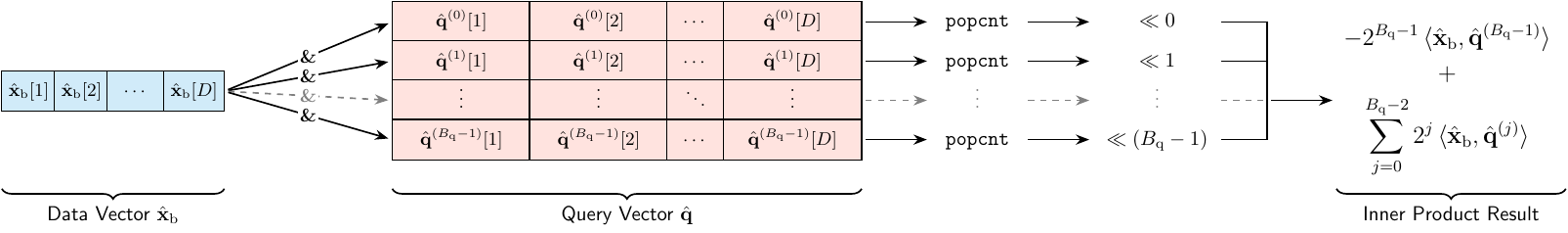}  
    \caption{Bitwise Inner Product Computation. $\hat{\mathbf{x}}_\mathrm{b}[i]$ denotes the $i$th dimension (bit) of the data vector $\hat{\mathbf{x}}_\mathrm{b}$, and $\hat{\mathbf{q}}^{(j)}[i]$ denotes the $j$th bit of the query vector's $i$th dimension $\hat{\mathbf{q}}[i]$. }
    \label{fig:bitwise_ip}
\end{figure*}

\smallskip
{\chengc \noindent\textbf{(1) Lookup Table based Strategy.}}
{\chengcr We precompute} query-dependent lookup tables (LUTs) and reuse them across all data vectors when evaluating the 1-bit estimated distance.
Different from GPU implementations of IVF-PQ (e.g., cuVS~\cite{cuvs_github}) that build LUTs over PQ codewords (i.e., per PQ dimension/subspace), we construct LUTs that aggregate inner-product contributions over multiple dimensions of the 1-bit RaBitQ code.
Formally, we partition the dimensions into groups of $U$ bits and construct multiple lookup tables accordingly. Let $m = D/U$ be the number of partitioned blocks, and 
\[
\mathbf{\hat{x}}_\mathrm{b}^{(j)}, \,\mathbf{q'}^{(j)} \in \mathbb{R}^{U}, \qquad j = 1,\ldots,m 
\]
be the subvectors corresponding to the $j$-th block {\cheng of the 1-bit code $\mathbf{\hat{x}}_\mathrm{b}$ and the query vector (after rotation) $\mathbf{q'}$}, {\chengc respectively}. For each block, we precompute a lookup table (LUT) that stores the inner products between $\mathbf{q'}^{(j)}$ and all 
possible $2^{U}$ binary patterns and store the values. {\chengc Specifically,} given the $j$th block, we define the corresponding LUT as follows:
\begin{equation}
L_j : \{0,1\}^{U} \to \mathbb{R}
\end{equation}
\begin{equation}
L_j(\mathbf{b}) = \langle \mathbf{q'}^{(j)}, \mathbf{b} \rangle
       = \sum_{i=1}^{U} {\cheng \mathbf{q}'}[(j-1){U}+i]\cdot \mathbf{b}[i] \ , 
\  \mathbf{b} \in \{0,1\}^{U}
\end{equation}
Given a binary code vector $\mathbf{\hat{x}}_\mathrm{b}$, 
the inner product $\langle \mathbf{\hat{x}}_\mathrm{b},  \mathbf{q'} \rangle $ can then be computed by summing the LUT 
entries corresponding to each block:

\begin{equation}
\langle \mathbf{\hat{x}}_\mathrm{b},  \mathbf{q'} \rangle 
= \sum_{j=1}^{m} \langle \mathbf{\hat{x}}_\mathrm{b}^{(j)},  \mathbf{q'}^{(j)}\rangle
= \sum_{j=1}^{m} L_j\big(\mathbf{\hat{x}}_\mathrm{b}^{(j)}\big).
\end{equation}
In practice, each binary block $\mathbf{\hat{x}}_\mathrm{b}^{(j)}$ is interpreted as an integer index 
$k \in \{0,\ldots,2^{U}-1\}$, allowing the inner product computation to be 
performed via $m$ table lookups followed by a summation. 

According to {\jifan Section \ref{subsec:rabitq}}, in IVF-RaBitQ, the inner product between $\mathbf{\hat{x}}_\mathrm{b}$ and $\mathbf{q'}$ has no relationship with which cluster $\mathbf{\hat{x}}_\mathrm{b}$ belongs to. That is, to compute the 1-bit estimated distance between a query vector and different data vectors, the LUTs only need to be computed once and can be reused across all data vectors. For each query, we have {\chengc $D/U$} LUTs, where each LUT uses the {\chengc $U$} bits as entry and has a size of \texttt{sizeof(lut\_type)} $ \times \, 2^{U}$ bytes. To enable high-throughput execution, an ideal case should be that LUTs from different queries can reside in an SM’s shared memory (e.g., up to 164 KB on Ampere and 228 KB on Hopper-class GPUs) at the same time, so that multiple queries can be processed concurrently on a single SM. Thus, we set $U$ to 4 to achieve the best performance. For example, if we use FP16 to store inner product results in LUT, for a 1,024-dimensional query, we only need {
\jifanb $1024/4\times2^4\times2 \mathrm{B}= 8 \mathrm{KB}$} to store all LUTs. 
Once the LUTs are loaded into the GPU's shared memory, they can be shared with hundreds of threads within a thread block with low access latency, enabling parallel inner-product evaluation for large batches of 1-bit RaBitQ codes.

\smallskip
{\chengc \noindent\textbf{(2) Bitwise Decomposition based Strategy.}}
While the LUT-based approach is effective, it relies on query-dependent tables residing in shared memory, which can limit on-chip residency and parallelism for high-dimensional vectors or GPUs with relatively small shared memory.
To complement this memory-centric design, we revisit the inner-product computation itself to develop a compute-centric scheme that minimizes shared-memory usage and better exploits GPU bitwise instructions.


{\cheng The classical inner-product computation processes vectors dimension by dimension. On GPUs, however, there is no native instruction that multiplies a packed bit string directly with floating-point values. As a result, a straightforward implementation must first extract each 1-bit entry from the bit string $\hat{\mathbf{x}}_\mathrm{b}$ (at least at byte granularity) before performing the multiplications.
Inspired by the bitwise decomposition in RaBitQ~\cite{10.1145/3654970}, we instead adopt a \emph{bitwise} inner-product computation that contrasts with the classical dimension-wise approach. The key idea is to reorganize the computation so that each instruction processes multiple dimensions at once using bit operations. To enable this, we first quantize each rotated query into $B_\mathrm{q}$-bit signed integers $\hat{\mathbf{q}}$. Based on Section~\ref{subsec:rabitq}, this quantization is performed once per query and reused across all probed clusters, contributing negligible overhead to the overall search pipeline. With the quantized query, 
the target inner product $\langle \hat{\mathbf{x}}_\mathrm{b}, {\chengc \mathbf{q}'} \rangle$ can be rewritten as:}

\begin{equation}
\langle \hat{\mathbf{x}}_\mathrm{b}, \mathbf{q}' \rangle
\;\approx\;
\Delta_\mathrm{q} \cdot \langle \hat{\mathbf{x}}_\mathrm{b}, \hat{\mathbf{q}} \rangle .
\end{equation}
{\cheng where $\Delta_\mathrm{q}$ is {\jifan the query quantization scale (step size) that maps the integer-domain inner product back to {\jifanb its} original scale {\jifanb (e.g., floating-point)} }}. 


{\chengc Let $\hat{\mathbf{q}}[i]$ be the $i$th dimension of $\hat{\mathbf{q}}$ where $1\le i\le D$ and} $\hat{\mathbf{q}}^{(j)}[i] \in \{0,1\}$ be the $j$th bit of $\hat{\mathbf{q}}[i]$ where $0 \le j < B_\mathrm{q}$ {\chengc (please refer to presentation of $\hat{\mathbf{q}}$ in Figure~\ref{fig:bitwise_ip} for illustration)}. The bitwise decomposition can be represented using the following equation:

\begin{align}
\langle \hat{\mathbf{x}}_\mathrm{b}, \hat{\mathbf{q}} \rangle
&= \sum_{i=1}^{D} \hat{\mathbf{x}}_\mathrm{b}[i] \cdot \hat{\mathbf{q}}[i] \\
&= \sum_{i=1}^{D} \hat{\mathbf{x}}_\mathrm{b}[i] \cdot
\left(
-2^{B_\mathrm{q}-1}\,\hat{\mathbf{q}}^{(B_\mathrm{q}-1)}[i]
+ \sum_{j=0}^{B_\mathrm{q}-2} 2^j\,\hat{\mathbf{q}}^{(j)}[i]
\right)
\label{eq:inner_product_expand} \\
&= -2^{B_\mathrm{q}-1}\,\langle \hat{\mathbf{x}}_\mathrm{b}, \hat{\mathbf{q}}^{(B_\mathrm{q}-1)} \rangle
 + \sum_{j=0}^{B_\mathrm{q}-2} 2^j\,\langle \hat{\mathbf{x}}_\mathrm{b}, \hat{\mathbf{q}}^{(j)} \rangle
\label{eq:inner_product_bitwise}
\end{align}

This equation depicts that the inner product can be viewed as a weighted sum of inner products between binary vectors $\mathbf{\hat{x}}_\mathrm{b}$ and $\mathbf{\hat{q}}^{(j)}$. As shown in Figure~\ref{fig:bitwise_ip}, the inner product between two binary vectors can be efficiently computed via bitwise \texttt{AND} {\chengd ($\&$)} and population count (\texttt{POPCNT}) operations. Since GPU registers are 32 bits wide, we can process 32 dimensions within a single operation, which provides an advantage over classical dimension-wise inner-product computation. Once $\langle \hat{\mathbf{x}}_\mathrm{b}, \hat{\mathbf{q}}^{(j)} \rangle$ is computed, the final weighted sum can be obtained through several shift and add operations. {\jifanrevision In this way, the bitwise method converts compression into computation: the compact 1-bit representation is not merely a storage format, but also the format in which the dominant inner product is evaluated.} Compared to the LUT-based approach, the bitwise inner product trades additional integer arithmetic for a substantially smaller shared-memory footprint.
On GPU, only the quantized query representation needs to be stored in shared memory (e.g., 0.5\,KB for a 1,024-dimensional query with 4-bit quantization), making this approach particularly suitable for high-dimensional settings or scenarios where shared memory is the limiting resource.

\subsubsection{Fused Cluster-local Search Kernel}
\label{subsec:fused_cluster_search_kernel}
Building on the optimized inner-product computation, we next design a fused cluster-local search kernel to efficiently execute the two-stage distance estimation on the GPU.
%
{\chengc Recall in Section~\ref{subsec:cluster_local_search} that the cluster-local search procedure for a $(\text{query}, \text{cluster})$ pair (which finds the $K$ nearest neighbors of the query within the cluster) consists of four logical steps, namely (1) 1-bit estimated distance computation, (2) candidate selection, (3) refined estimated distance computation, and (4) in-block top-$K$ selection}. A straightforward approach is to implement each of {\chengc these} steps as a separate kernel.
However, due to data dependencies between consecutive steps, this design requires launching multiple kernels sequentially, where each kernel loads its inputs from global memory and writes intermediate results back to global memory.
We note that shared memory is scoped to a single thread block and cannot be shared across blocks in different kernels, which means that it is necessary to store intermediate results in global memory between stages.
Moreover, when workload imbalance arises across thread blocks, later kernels must wait for the completion of earlier ones, leading to long-tail effects and underutilization of streaming multiprocessors (SMs).


To address these inefficiencies, we design a fused cluster-local search kernel that integrates all four logical steps into a single kernel execution.
Rather than launching multiple kernels for different stages, the fused design propagates intermediate results across stages using on-chip memory, enabling block-level synchronization instead of kernel-level synchronization.
This design significantly reduces kernel launch overhead, global-memory traffic, and synchronization costs, especially under workload imbalance.

\smallskip
\noindent\textbf{Kernel organization.}
The fused kernel adopts a block-centric execution model,  where each block takes a $(\text{query}, \text{cluster})$ pair as an input and returns the cluster-local $K$-nearest neighbours for the query.
Upon launch, a block first loads the query-dependent representation into shared memory. Depending on the selected inner-product {\cheng computation method} (Section~\ref{subsec:ip_gpu}), this representation is either the precomputed lookup tables (LUTs) or the quantized query vector used for bitwise inner-product evaluation. After that, threads in the block collaboratively scan the data vectors in the cluster and compute lower bounds of the 1-bit estimated distances. Each thread processes one or more vectors and evaluates the inner product between the query representation and the 1-bit RaBitQ codes. Based on the inner product results, the 1-bit estimated distance lower bounds are computed and compared against a query-specific threshold  (the current $K$-th largest distance among the candidates found so far across clusters).
Vectors whose lower bounds exceed the threshold are pruned early, while the remaining vectors are selected as candidates for further refinement.
Then, the fused kernel performs distance refinement for the surviving candidates.
Since the number of candidates typically constitutes a small fraction of the cluster size, we assign groups of threads within a warp to cooperatively process each candidate. Specifically, threads within a warp jointly compute the inner product between the query and the ex-codes of a candidate, which helps amortize the cost of accessing additional code bits and achieves better workload balance for cases where the number of candidates is small. Using the refined inner-product results together with the associated auxiliary factors, the block computes the refined estimated distance for these candidates. Finally, each block {\cheng conducts} a cluster-local top-$K$ selection using shared memory. After all vectors in the cluster have been processed, the block writes the cluster-local top-$K$ results back to global memory for subsequent merging across clusters, and updates the query-specific threshold atomically.

\smallskip
\noindent\textbf{Shared-memory management via stage-aware reuse.}
A key challenge of kernel fusion is managing shared-memory usage across stages while preserving sufficient SM residency. Unlike CPU programs that can allocate and free memory dynamically, the shared-memory footprint of a GPU kernel must be determined at launch time, as it directly constrains the number of thread blocks that can concurrently reside on an SM. Excessive shared-memory allocation can therefore reduce occupancy and limit overall throughput. We observe that the shared-memory requirements of different stages in the fused kernel do not fully overlap in time. Specifically, the memory used to store the query-dependent representation (i.e., lookup tables or quantized query vectors) is only needed during the 1-bit distance estimation stage, while later stages require shared memory primarily for storing intermediate inner-product results, candidate metadata, and top-$K$ buffers.
Exploiting this property, we reuse shared-memory regions across stages whenever their lifetimes do not overlap: (1) the shared-memory region initially used to store the LUTs or quantized queries is reused to store intermediate inner-product results during distance refinement; (2) the shared memory allocated across the first three stages is repurposed as a sorting buffer for maintaining the cluster-local top-$K$ results in the final stage. By carefully coordinating these reuse patterns, the fused kernel achieves a substantially smaller peak shared-memory footprint than a naive allocation that reserves separate buffers for all stages.

\begin{figure*}[!thbp] 
    \centering
    \includegraphics[width=0.98\linewidth]{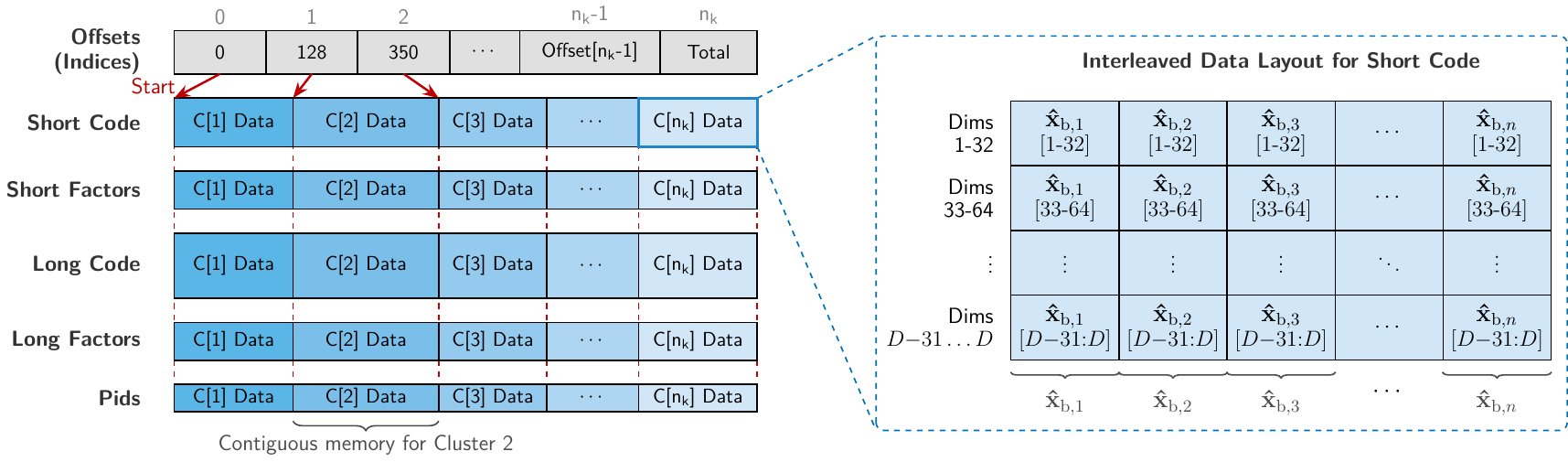}  
    \caption{{\jifanrevision GPU-oriented IVF-RaBitQ index layout.}}
    \label{fig:data_layout}
\end{figure*}

\subsubsection{GPU-Oriented Index Layout}
\label{subsec:optimized_GPU_data_layout}

To support efficient execution of the fused kernel, we co-design a GPU-oriented index layout that aligns data organization with cluster-local access patterns.
IVF-RaBitQ follows a flattened IVF organization inspired by Compressed Sparse Row (CSR)~\cite{10.1145/1583991.1584053} techniques, where vectors belonging to the same cluster are stored contiguously and indexed via an offset array. Building on this baseline organization, we further reorganize RaBitQ codes and auxiliary data to improve data locality, memory coalescing, and cache efficiency during GPU execution.
This layout is co-designed with the fused kernel to enable high-throughput search on modern GPUs.

\smallskip
\noindent\textbf{Stage-aware data decomposition.}
To match the multi-stage execution of the fused kernel, the data associated with each cluster is decomposed into multiple contiguous arrays according to their usage patterns.
Specifically, we separate the RaBitQ representation into short codes {\jifanrevision (the 1-bit RaBitQ codes used in the filtering stage)} and long codes {\jifanrevision (the remaining ex-codes used in the refinement stage)}.
Correspondingly, auxiliary factors required for distance estimation are split into short-factor and long-factor arrays, {\cheng used for computing $1$-bit distances and refined distances, respectively}.
In addition, a {\jifanb \texttt{Pid}} array stores the unique identifier of each vector within the cluster.
{\jifanrevision As illustrated in the left part of Figure~\ref{fig:data_layout}, IVF-RaBitQ stores each component of the index in a separate flattened array. The offset array contains $n_k+1$ entries, where two consecutive offsets delimit the segment of one cluster. These offsets are shared by all arrays, so the short codes, short factors, long codes, long factors, and vector identifiers (\texttt{Pid}) belonging to the same cluster can be located consistently.}
This stage-aware decomposition improves data locality and avoids loading unused data in early stages of the fused kernel.

\smallskip
\noindent\textbf{Interleaved layout for 1-bit codes.}
Efficient inner-product computation in the filtering stage requires high memory bandwidth when accessing the 1-bit RaBitQ codes.
On GPUs, global memory accesses are serviced via aligned memory transactions (e.g., 32, 64, or 128 bytes), and accesses from consecutive threads in a warp can be coalesced when they target adjacent addresses.
To exploit this behavior, we design an interleaved layout for the 1-bit codes.
{\jifanrevision The right part of Figure~\ref{fig:data_layout} further shows how the short-code segment of a cluster is organized. For a cluster containing $n$ vectors of dimension $D$, the 1-bit code of each vector is packed every 32 dimensions (e.g., dimensions 1--32, 33--64) into a 32-bit unsigned integer. Instead of storing all packed words of one vector consecutively, we store the packed words of all $n$ vectors for the same 32-dimensional group contiguously before moving to the next group.}
With this layout, threads within a warp access adjacent 32-bit words when processing the same dimension group across different vectors.
As a result, memory accesses are coalesced into 128-byte transactions and fit within a single L1 cache line, enabling efficient utilization of GPU memory bandwidth during inner-product computation.

\section{Experiments}
\label{sec:exp}

\subsection{Experimental Setup}
\label{subsec:exp_setup}

\smallskip \noindent \textbf{Experimental Platform.} The GPU-based experiments are conducted on a machine provided by Amazon Web Services \cite{aws_ec2_g6e}. The machine is equipped with a single NVIDIA L40S GPU with 48 GB of device memory, and an AMD EPYC 7R13 processor providing 8 virtual CPU cores {\jifanrevision with 64GB main memory}. The NVIDIA driver version is 590.48.01 with CUDA 13.1. The CPU-based experiments are performed on a machine equipped with {\jifanrevision 2} Intel Xeon Gold 6418H processors and 1 TB of  {\jifanrevision
DDR5-4800 registered ECC memory, organized as 16 $\times$ 64 GB dual-rank DIMMs across two NUMA nodes}. Both machines run Ubuntu 22.04 LTS (x86\_64).

\smallskip \noindent \textbf{Datasets.} We use {\jifanrevision nine} modern real-world high-dimensional vector datasets to evaluate the performance, as detailed in Table~\ref{tab:datasets}. These datasets are taken from the popular ANNS benchmarks and reflect real-world ANNS search scenarios~\cite{jaasaari2025vibevectorindexbenchmark,  gqr_datasets, openai2024embedding, rapids2024wikiall}.
They cover image ({\jifanrevision \textit{SIFT}~\cite{5432202}, \textit{DEEP}~\cite{7780595, 8100152}, including DEEP-1M with 256 dimensions and DEEP-150M, a 150M-vector cut from DEEP-1B with 96 dimensions,} \textit{ImageNet}~\cite{5206848}, and \textit{GIST}~\cite{5432202}), code (\textit{CodeSearchNet}~\cite{husain2019codesearchnet}), and text (\textit{OpenAI-3072-1M}, \textit{OpenAI-1536-5M}\cite{openai2024embedding} and \textit{Wiki-all}~\cite{rapids2024wikiall}). 
For each dataset (except for Wiki-all {\jifanrevision and SIFT}), we exclude 10,000 vectors from the dataset as query vectors, and use the rest of them as base data vectors for 
index construction. 
For Wiki-all {\jifanrevision and SIFT}, {\cheng the dataset includes 10,000 query vectors already.}

\begin{table}[!t]
\centering
\caption{Datasets used in evaluation. $N$ is the total number of database vectors and $D$ is the number of dimensions.}
\label{tab:datasets}
\begin{tabular}{l c c c c}
\toprule
\textbf{Name} & \textbf{Type} & $\boldsymbol{N}$ & $\boldsymbol{D}$ & \textbf{Size (GB)}  \\
\midrule
{\jifanrevision SIFT-1M}   & {\jifanrevision Image} & {\jifanrevision 1,000,000}  & {\jifanrevision 128}  & {\jifanrevision 0.52}  \\
{\jifanrevision DEEP-1M}   & {\jifanrevision Image} & {\jifanrevision 1,000,000}  & {\jifanrevision 256}   & {\jifanrevision 1.02}  \\
ImageNet   & Image & 1,281,167  & 512  & 2.62  \\
CodeSearchNet  & Code  & 1,374,067  & 768  & 4.22  \\
GIST                           & Image & 1,000,000  & 960  & 3.84   \\
OpenAI-3072-1M  & Text  & 1,000,000  & 3072 & 12.29  \\
OpenAI-1536-5M          & Text  & 5,000,000  & 1536 & 30.78 \\
Wiki-all                  & Text  & 10,000,000 & 768  & 30.75  \\
{\jifanrevision DEEP-150M}                  & {\jifanrevision Image} & {\jifanrevision 150,000,000} & {\jifanrevision 96}  & {\jifanrevision 57.60} \\
\bottomrule
\end{tabular}
\end{table}

\begin{figure*}[!thbp] 

    \centering
    \includegraphics[width=1\linewidth]{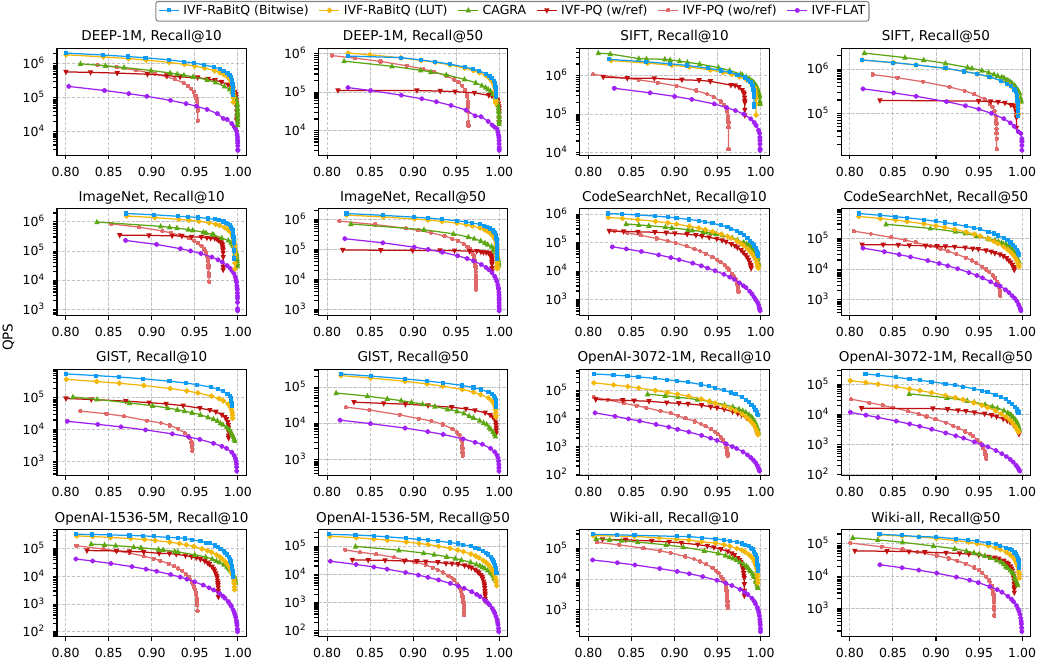}  
    \caption{{
Time--accuracy trade-off of ANN search on representative datasets (log-scale).
\emph{Bitwise} and \emph{LUT} denote the two GPU inner-product methods used by IVF-RaBitQ,
while \emph{w/ refine} and \emph{w/o refine} indicate whether distance refinement is enabled for IVF-PQ.
}}
    \label{fig:recall-qps}
\end{figure*}

\smallskip\noindent \textbf{{\chengd Baselines}.} 
We compare our method against several modern ANN methods. For GPU-based methods, we first choose {\chengc the widely used} cluster-based methods, IVF-Flat and IVF-PQ, implemented in the cuVS Library~\cite{cuvs_github}. These methods are built on the inverted file (IVF) {\chengc index}.
\textit{IVF-Flat}~\cite{Feher2023IVFFlat} stores all database vectors in uncompressed form.
\textit{IVF-PQ}~\cite{Chirkin2024IVFPQ} applies product quantization~\cite{5432202} to compress database vectors into compact codes. 
It {\cheng provides two options, namely one with refinement (named IVF-PQ (w/ref)) and the other without refinement (named IVF-PQ (wo/ref)).
IVF-PQ (w/ref)} 
re-evaluates shortlisted candidates {\cheng produced based on the PQ codes} using raw data vectors to improve recall. 
Apart from {\cheng GPU-based methods that are based on clustering}, we also include \textit{CAGRA}~\cite{10597683}, a state-of-the-art graph-based GPU ANNS algorithm that constructs and searches fixed-degree proximity graphs fully in parallel. For CPU-based methods, we include \textit{RaBitQLib}~\cite{gao2025the} (which implements {\cheng IVF-RaBitQ} for ANNS with efficient SIMD instructions).
{\jifanb Among these competitors, IVF-PQ and RaBitQLib provide the option for vector quantization. For IVF-PQ (w/ref) and IVF-PQ (wo/ref), we choose the best index settings (e.g., quantization dimensions, quantization bits, and number of clusters) for searching at Recall=0.95 to evaluate the time-accuracy tradeoff. For IVF-RaBitQ and RaBitQLib, we quantize each dimension of vectors to 8 bits unless mentioned otherwise. For RaBitQ quantization on GPU, we empirically set $N_{\mathrm{coarse}}$ and $N_{\mathrm{fine}}$ to 64 and 32, respectively.}



\smallskip \noindent  \textbf{Benchmark.} All experiments with GPU are conducted using \textit{cuVS Bench}~\cite{rapids_cuvs_bench}, an open-source benchmarking suite developed by NVIDIA for evaluating approximate nearest neighbor (ANN) algorithms on GPUs. It enables fair comparison across methods by enforcing consistent data loading, query batching, and performance measurement under GPU-resident settings. 
Unless mentioned otherwise, we use $batchsize=10^4$ for the benchmark.

\subsection{Experimental Results}

\subsubsection{GPU ANNS Search Performance}
\label{subsec:gpu_search_perf}

\begin{figure*}[!thbp]
  \centering
  \includegraphics[width=\textwidth]{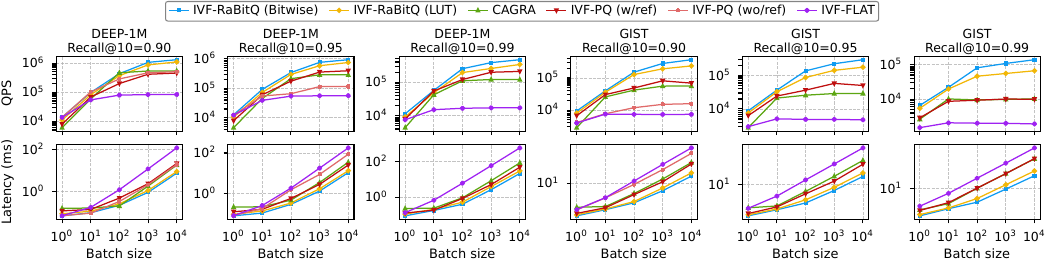}
  \caption{{\jifanrevision Throughput and Latency scaling with batch size on DEEP-1M and GIST dataset.}}
  \label{fig:batch-size-qps-latency}

\end{figure*}

\begin{figure*}[!thbp]
  \centering
  \includegraphics[width=\textwidth]{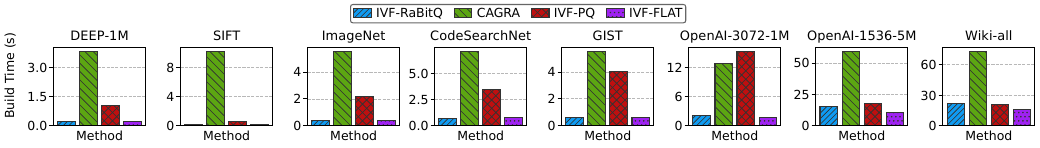}
  \caption{Index build time across different datasets. Each subplot compares build time across methods.}
  \label{fig:build-time-grid}

\end{figure*}

\begin{figure*}[!thbp]
  \centering
  \includegraphics[width=\textwidth]{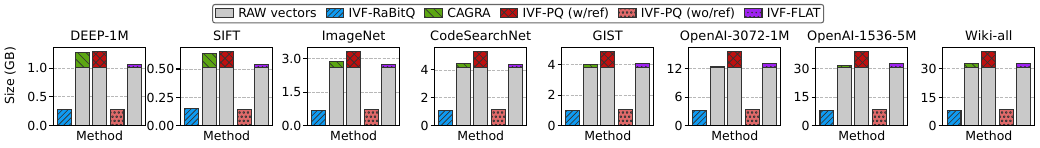}
  \caption{Storage requirement of different indexing methods. CAGRA and IVF-PQ (w/ref) require RAW vectors for refinement, thus {\chengcr the storage of RAW vectors} is included.}
  \label{fig:storage-requirement}

\end{figure*}

In this subsection, we evaluate the GPU-based ANN search performance of IVF-RaBitQ against other representative baselines. 
For IVF-RaBitQ, we report two different GPU search modes: \emph{IVF-RaBitQ (LUT)} and \emph{IVF-RaBitQ (Bitwise)}, which correspond to two different inner product computation methods (recall Section~\ref{subsec:ip_gpu}) used in our fused kernel. 
Our competitors include IVF-PQ (with and without refinement), IVF-Flat, and CAGRA. 
{\jifanrevision All indexes are pre-loaded into GPU memory before the search starts.} 
Figure~\ref{fig:recall-qps} reports the time-accuracy trade-off for these methods on representative datasets.
The performance is evaluated in QPS (query-per-second), and the accuracy is measured by Recall@K (the ratio of true nearest neighbors found in the top-$K$ results returned by an ANNS algorithm
). Based on the results, we have the following observations. 
\ul{(1) Comparison among IVF-based methods.} Across all datasets, IVF-RaBitQ significantly outperforms IVF-PQ in terms of throughput, especially at higher recall targets. 
Although IVF-PQ with refinement (denoted as IVF-PQ (w/ref)) improves retrieval accuracy by introducing an additional refinement stage, {\jifanb it still incurs considerable computational and memory overhead when using raw vectors for reranking}. In contrast, IVF-RaBitQ attains a more favorable efficiency--accuracy balance without relying on reranking of raw vectors.
{\jifancamera To be specific, for Recall=0.95, IVF-RaBitQ has 2.4$\times$\textasciitilde58.3$\times$ speedup compared to {\chengcr IVF-PQ (wo/ref),}
and 1.4$\times$\textasciitilde10.0$\times$ speedup compared to IVF-PQ (w/ref).}
%
\ul{(2) Comparison with CAGRA.} IVF-RaBitQ demonstrates comparable or higher throughput than CAGRA across most datasets and recall ranges. IVF-RaBitQ benefits from compression while CAGRA works with uncompressed data. On average, IVF-RaBitQ (Bitwise) achieves a speed up of {\jifancamera 2.9$\times$}, {\jifancamera 3.3$\times$, 4.0$\times$} compared to CAGRA for recall={0.9, 0.95, 0.99}@10, respectively.
The performance advantage becomes more evident on higher recall, which highlights the advantage of cache reuse while probing many clusters for a large batch of queries.
{\jifanrevision We note that when the dimensionality is relatively low (e.g., SIFT with $D=128$), CAGRA {\chengr has a slightly better performance} than IVF-RaBitQ, which is attributed to a lower computational overhead of CAGRA's graph traversal at low {\chengrb dimensionality}.}
\ul{(3) Bitwise vs. LUT-based GPU kernels.}
We further investigate two inner product computation methods for IVF-RaBitQ: a bitwise inner-product kernel and a lookup-table (LUT) based kernel.
Both implementations achieve high performance across all datasets.
As dimensionality increases, the LUT-based kernel is bounded by shared-memory capacity, whereas the bitwise formulation remains lightweight and scalable {\cheng which is reflected by gap between the two methods on the OpenAI-3072-1M dataset}. We note that the relative performance of these two methods may vary based on the memory bandwidth and shared memory size of different GPUs.
\ul{(4) Scalability across dataset sizes and dimensionalities.}
Across datasets ranging from 1M to 10M vectors and dimensions from {\jifanrevision 256} to 3072, IVF-RaBitQ maintains robust performance and consistent throughput advantages.



{\chengr
\subsubsection{Scalability with Query Batch Size}
\label{subsec:batch_size_scalability}
We compare the methods by varying the query batch size. Figure~\ref{fig:batch-size-qps-latency} reports the throughput and latency as a function of batch size for target Recall@10 of 0.90, 0.95, and 0.99. In terms of throughput, IVF-RaBitQ is comparable to IVF-PQ (w/ref) and CAGRA at small batch sizes (1 and 10), while significantly outperforming both at larger batch sizes (100, 1{,}000 and 10{,}000), where GPU parallelism is more fully utilized. In terms of latency, all methods achieve sub-millisecond response times at small batch sizes. As the batch size increases, IVF-RaBitQ maintains substantially lower latency than the competing methods, demonstrating favorable scalability across workload intensities.}

\subsubsection{Index Build Time}
\label{subsec:index_build_time}
Figure~\ref{fig:build-time-grid} shows the index build time for the indexes used in Section~\ref{subsec:gpu_search_perf}. For all IVF-based methods, we use {\jifanb one-tenth of the dataset} vectors to train the clusters {\jifanb and align the number of clusters to be trained. Also, we align the IVF-PQ and IVF-RaBitQ to use the same compression ratio for quantization (i.e., 8 bits per dimension).}   We {\cheng observe} that IVF-RaBitQ {\jifanb has a very short} index build time across all evaluated datasets. When compared with CAGRA, the proposed method attains an average speedup of {\jifancamera $14.7\times$}. 
{\cheng This is because CAGRA} needs to spend more time on its KNN graph building and graph optimization procedures. 
This overhead becomes increasingly dominant as the dataset scale and dimensionality grow.
Compared with other IVF-based methods, IVF-RaBitQ {\jifanb is overall} faster than IVF-PQ {\cheng (with a {\jifanrevision $4.2\times$} speedup on average)} and {\jifanb is only suboptimal to} IVF-Flat. 


\subsubsection{Storage Footprint}
\label{subsec:storage_requirement}
We analyze the storage footprint of different ANN indices, which is the total amount of data that is stored to support search, as shown in Figure~\ref{fig:storage-requirement}. We note that IVF-PQ (wo /ref) uses the same compression ratio (8 bits per dimension for a floating-point vector) as IVF-RaBitQ. CAGRA, IVF-Flat, and IVF-PQ with refinement need to store the raw data vectors to support exact or refined distance computation, which significantly increases their storage demand. Compared to these methods, IVF-RaBitQ only needs less than 25\% of their storage requirement to achieve a comparable recall. 
Compared with IVF-PQ (wo/ref), which 
does not require raw vectors, IVF-RaBitQ achieves a similar storage footprint while delivering substantially higher recall and query throughput. This advantage {\james could enable deploying} large-scale ANN systems within the limited memory budget of modern GPUs.

\subsubsection{Ablation Study for Kernel and Data Layout Co-design.}
\label{subsec:ablation_kernel_layout}
\begin{figure}[!t]
    \centering
    \includegraphics[width=\linewidth]{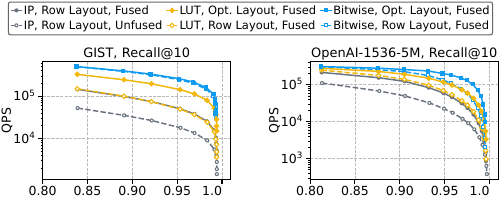}
    \caption{{\jifanrevision
Ablation study for kernel and data layout co-design.
}}
    \label{fig:ablation_study}
\end{figure}

{\chengr We conduct an ablation study to quantify the individual contributions of the three key design dimensions in IVF-RaBitQ: inner-product computation method, data layout, and kernel organization. As shown in Figure~\ref{fig:ablation_study}, we evaluate the following variants: (1)~three inner-product computation methods---na\"ive floating-point inner product (IP), lookup-table based (LUT), and bitwise decomposition based (Bitwise); (2)~two data layouts---a row-major na\"ive layout (Row Layout) and our optimized bit-major interleaved layout (Opt.\ Layout); and (3)~two kernel organizations---separate unfused kernels and our fused cluster-local search kernel. The results confirm that each design dimension contributes meaningfully to overall performance: the fully optimized LUT and Bitwise kernels achieve over {\jifancamera$5\times$} higher throughput than the na\"ive unfused baseline at recall levels above 0.95. Notably, the benefit of the optimized data layout is more pronounced on high-dimensional datasets, where global-memory traffic is more intensive and coalesced access patterns yield a larger reduction in memory stalls.}

\subsubsection{ANNS Search for Datasets Exceeding GPU Memory}
\label{sec:deep150m}
\begin{figure}[!t]
    \centering
    \includegraphics[width=\linewidth]{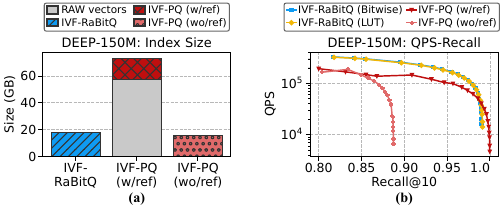}
    \caption{{\jifanrevision
ANNS search on DEEP-150M, where the raw vectors exceed the memory capacity of a single L40S GPU: (a) storage {\chengr footprint} of IVF-RaBitQ and IVF-PQ variants, and (b)
time-accuracy tradeoff.
}
}
    \label{fig:deep150m_oogpu}
\end{figure}

{\jifanrevision
We evaluate IVF-RaBitQ on DEEP-150M, whose raw base vectors occupy about {\chengr 57.6\,GB}, which is larger than the 48\,GB device memory of a single NVIDIA L40S GPU, yet within the 64\,GB host memory capacity. Figure~\ref{fig:deep150m_oogpu} reports both the storage footprint and the time--accuracy trade-off for all applicable methods. CAGRA and IVF-Flat cannot handle this dataset due to their large runtime memory requirements. {\chengcr Note that both of them use raw vectors {\jifancamera on GPU} for {\jifancamera searching}.} IVF-RaBitQ requires only about {\chengr 18.0 GB} for its index and does not require raw vectors during search. In contrast, IVF-PQ with refinement must retain both the compressed index and the raw vectors, resulting in a total footprint of {\chengr 73.3\,GB}. For search performance, IVF-RaBitQ achieves $1.58\times$ and $1.70\times$ higher QPS than IVF-PQ (w/ref) at Recall$\,\geq 0.90$ and $\geq 0.95$, respectively, while IVF-PQ (wo/ref) cannot reach these recall levels at all.
}


\subsubsection{Quantization Accuracy and Runtime.}
\label{subsec:quant_accuracy_runtime}
\begin{figure}[!thbp]
    \centering
    \includegraphics[width=\linewidth]{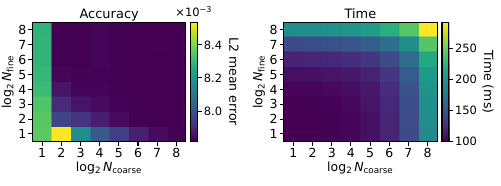}
    \caption{{\jifanrevision
Quantization accuracy and runtime {\chengr of the GPU-based RaBitQ quantization} at 8 bits on the GIST dataset.
}}
    \label{fig:gist_8bit_accuracy_time}
\end{figure}

{\chengr We evaluate the trade-off between quantization accuracy and runtime by varying $N_{\mathrm{coarse}}$ and $N_{\mathrm{fine}}$ in our two-phase parallel grid search method.
In this experiment, we quantize the normalized vectors without building an index structure, and report the mean $\ell_2$ quantization error alongside the wall-clock quantization time. Figure~\ref{fig:gist_8bit_accuracy_time} shows that, on the GIST dataset with 8-bit quantization, setting both $N_{\mathrm{coarse}}$ and $N_{\mathrm{fine}}$ to as few as 16 already yields a mean quantization error below 0.008. We adopt slightly conservative defaults of $N_{\mathrm{coarse}} = 64$ and $N_{\mathrm{fine}} = 32$ to ensure high quantization accuracy with negligible additional overhead.}

\subsubsection{Comparison with CPU-based {\chengb Counterpart}}
\label{subsec:cpu_counterpart}

\begin{figure}[!t]
    \centering
    \includegraphics[width=\linewidth]{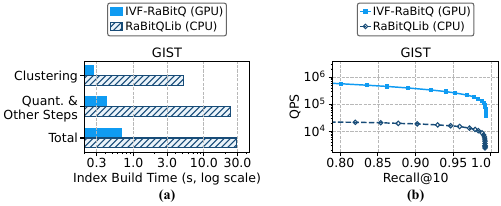}
    \caption{
{\jifanrevision (a) Index build time breakdown, (b) Time-accuracy trade-off of IVF-RaBitQ and its CPU counterpart (RaBitQLib) on GIST dataset. 
The GPU implementation runs on an NVIDIA L40S, and the CPU counterpart runs on an Intel Xeon Gold 6418H with 48 threads.}
}
    \label{fig:qps_recall_bits_gpu_cpu}
\end{figure}

We further compare IVF-RaBitQ {\chengb (GPU) with the bitwise inner product computation method 
and its counterpart on CPU, for which we adopt the implementation from the}
RaBitQLib~\cite{gao2025the}. {\jifanrevision The CPU counterpart is bound to one physical
CPU socket (utilizing 48 threads) on a fixed NUMA node, with memory allocated from the same NUMA node to avoid cross-socket migration and remote-NUMA effects. Figure~\ref{fig:qps_recall_bits_gpu_cpu} (a) shows the build time comparison between RaBitQLib and IVF-RaBitQ (GPU) on GIST dataset, where IVF-RaBitQ (GPU) is 42.3$\times$ faster.} Figure~\ref{fig:qps_recall_bits_gpu_cpu} (b) reports the time-accuracy trade-off curves 
{\chengr on the GIST dataset.}
At Recall@10 = 0.9, 0.95, and 0.99, IVF-RaBitQ {\chengb (GPU)} achieves QPS speedups of {\jifancamera $19.8\times$, $16.8\times$, and $15.1\times$}, {\chengd respectively} over {\chengb its CPU counterpart}. 




\section{Related Work}
\label{sec:related}



\noindent\textbf{Graph-Based ANNS on GPU.}
Graph-based methods typically build a proximity graph 
and answer queries by iteratively expanding a candidate frontier from one or a few entry points, while pruning candidates based on distance comparisons. 
{\jifanrevision The graph-based ANNS search algorithms work cooperatively with the graph index, and recent GPU designs mainly focus on accelerating graph traversal and candidate management (e.g., priority queues, visited sets, and hash-table-based filtering) under massive parallelism.}
 Graph-based GPU Nearest Neighbor (GGNN) \cite{9739943} proposes a parallel, bottom-up hierarchical construction algorithm that recursively merges smaller, independent sub-graphs into a final global index. 
GGraphCon \cite{9835618} is a GPU-based graph construction framework which uses a divide-and-conquer framework that partitions the graph building into different small tasks, and a sort-based strategy to build high-quality graphs in parallel. 
 {\jifanrevision CAGRA~\cite{10597683} uses a fixed-degree graph and rank-based edge reordering to improve GPU load balance and graph reachability, while its search kernel employs software warp splitting and forgettable hash-table management to reduce visited-set overhead. BANG~\cite{11045134} targets billion-scale graph search beyond GPU memory by keeping a compressed representation of vectors on the GPU while accessing the graph index from host memory, thereby optimizing the CPU--GPU memory layout rather than changing the graph traversal paradigm. }
{\chengc A recent work, concurrent with ours, 
{\chengd develops a graph-based method called Jasper for ANNS on GPU,
which uses RaBitQ for vector quantization and supports data updates~\cite{mccoy2026gpuacceleratedannsquantizedspeed}.}
We note that this work computes the rescaling factor on the CPU, while performing the rescaling and rounding on the GPU - an execution split that distinguishes it from our approach.}
{\jifanrevision
Tagore~\cite{10.1145/3769825} focuses on graph index construction and formulates graph-pruning procedures as Collect-Filter-Store kernels, enabling parallel edge pruning for refinement-based graph indexes. }
{\jifanrevision
{\chengrb These methods} accelerate complex graph construction and irregular, query-dependent graph traversal, where performance is dominated by frontier expansion, duplicate detection, neighbor-list access, and candidate pruning. In contrast, IVF-RaBitQ follows a cluster-based scan over selected inverted lists and avoids maintaining graph frontiers or visited sets during search.}
\smallskip
\noindent\textbf{Cluster-based ANNS on GPU.}
While most graph-based methods focus on maximizing the search throughput, the cluster-based methods usually deliver faster index builds, lower memory usage, and a strong balance between speed and accuracy \cite{li2025ivf_vs_hnsw}. In 2016,  Wieschollek et al. proposed
a two-level product and vector quantization tree that reduces the number of vector comparisons for ANNS
tree traversal, which firstly demonstrates
GPU performance is superior to CPU performance in large-scale ANN problems \cite{7780592}. Later, FAISS \cite{8733051} develops a high-performance top-$K$ selection algorithm on a GPU and integrates it with IVF-PQ for billion-scale ANNS search on multiple GPUs. Chen et al. proposed a hierarchical
inverted index structure generated by line quantization to enable more fine-grained search in IVF~\cite{CHEN2019295}. Recently, Nvidia cuVS \cite{Chirkin2024IVFPQ} further optimizes LUT residency in shared memory (via low-precision LUT types and kernel-path switching), and employs fused/early-stop kernels for IVF-PQ, making it a state-of-the-art implementation for IVF-PQ on GPU. Rummy \cite{10.5555/3691825.3691827} and FusionANNS \cite{10.5555/3724648.3724659} both focus on the scenario when vector data is larger than the GPU global memory, where Rummy is based on pure IVF and FusionANNS further extends it with Product Quantization and a navigation graph.
{\jifanrevision JUNO~\cite{10.1145/3620665.3640360} observes that PQ-based ANN search may perform unnecessary pairwise distance calculations and accumulations, and therefore introduces a sparsity- and locality-aware search algorithm mapped to ray-tracing cores and tensor cores.}
{\jifanrevision In contrast to these GPU cluster-based solutions that primarily build on IVF-Flat/IVF-PQ (and their codebook-based distance estimators), and optimize either PQ-LUT evaluation, cross-device data placement, or re-ranking I/O, our IVF-RaBitQ integrates RaBitQ and introduces GPU-native quantization, inner-product computation, and fused-kernel designs tailored to this method.}

\section{Conclusion}
\label{sec:conclusion}

We present IVF-RaBitQ (GPU), an end-to-end GPU-native ANNS framework that integrates IVF partitioning with 
{\cheng RaBitQ}
quantization. IVF-RaBitQ enables fast index build, high-throughput search, high recall, and compact storage by combining a scalable GPU quantization pipeline, GPU-native inner-product computation, a fused cluster-local search kernel, and a GPU-oriented index layout.
{\chengd We further integrate IVF-RaBitQ into NVIDIA \textbf{cuVS} and the \textbf{cuVS bench} pipeline to support reproducible evaluation and fair comparisons with existing GPU baselines.}
Experiments on {\cheng various} datasets 
show that IVF-RaBitQ offers a strong trade-off frontier across recall, throughput, build time, and memory footprint, especially in high-recall regimes.
{\jifanrevision In the future, we plan to explore extensions such as support for dynamic index updates and transactional operations.}

\begin{acks}
We sincerely appreciate the insightful discussions and constructive suggestions provided by Artem Chirkin, Akira Naruse, and Hiroyuki Ootomo at NVIDIA, as well as Yutong Gou and Yuexuan Xu at Nanyang Technological University.
This research is supported by the Ministry of Education, Singapore, under its Academic Research Fund (Tier 2 Award MOE-T2EP20224-0011 and Tier 1 Award (RG20/24)). Any opinions, findings and conclusions or recommendations expressed in this material are those of the author(s) and do not reflect the views of the Ministry of Education, Singapore.
\end{acks}

\balance

\bibliographystyle{ACM-Reference-Format}
\bibliography{sample}

\end{document}